\documentclass[aps,prl,twocolumn,showpacs,preprintnumbers,amsmath,amssymb,groupedaddress]{revtex4-1}
\usepackage{mathptm} 
\usepackage{dcolumn}                    
\usepackage{bm}                        
\usepackage{graphicx}
\usepackage{times}
\begin{document}
\newcommand {\be}{\begin{equation}}
\newcommand {\ee}{\end{equation}}
\newcommand {\bea}{\begin{eqnarray}}
\newcommand {\eea}{\end{eqnarray}}
\newcommand {\nn}{\nonumber}
\newcommand {\bb}{\bibitem}
\newcommand{\et}{{\,\it et al\,\,}}
\newcommand{\CuB}{CuBiS$_{2}$}
\newcommand{\BiTe}{Bi$_{2}$Te$_{3}$ }

\title{High three dimensional
thermoelectric performance from low dimensional bands}

\author{David Parker, Xin Chen  and David J. Singh}
\address{Materials Science and Technology Division, Oak Ridge National Laboratory, 1 Bethel Valley Rd., Oak Ridge, TN 37831-6056}

\date{\today}

\begin{abstract}
Reduced dimensionality has long been regarded as an important strategy for increasing thermoelectric performance, for example in superlattices and other engineered structures.  Here we point out and
illustrate by examples that three dimensional {\it bulk} materials can be made to behave as if they were two dimensional from the point of view of thermoelectric performance.  Implications
for the discovery of new practical thermoelectrics are discussed.

\end{abstract}
\pacs{}
\maketitle
\section{Introduction} 

Thermoelectric performance is quantified by the
figure of merit, $ZT=\sigma S^2 T/\kappa$,
where $\sigma$ is the electrical conductivity, $\kappa$ is
the thermal conductivity, $S$ is the thermopower (Seebeck
coefficient) and $T$ is the absolute temperature.
\cite{ioffe,wood}
There is no known thermodynamic or other fundamental
limitation on $ZT$, but finding high $ZT$ materials is
very challenging and only a handful of materials with
$ZT$ significantly higher than unity are known.
The difficulty is that finding high $ZT$ requires
finding a material that combines transport properties that
do not normally occur together. Here we focus on the combination
of high thermopower and high conductivity.

The low $T$ electrical conductivity of a metal or degenerate semiconductor
depends on the electronic states
and their scattering at the Fermi level, $E_F$, specifically
$\sigma\propto N(E_F)<v^2>\tau$, where $N$ is the density of 
states, $<v^2>$ is the average Fermi velocity for the current
direction, and $\tau$ is an inverse scattering rate.
\cite{ziman,jones}
At finite temperature the expressions are similar but
are integrated with the derivative of the Fermi function, which
amounts to a temperature broadening.
The conductivity therefore improves as one moves $E_F$ away from
the band edge, as in that case both the velocity and $N(E_F)$
increase.
The thermopower is different. At low $T$, $S(T)\propto T(d\sigma/dE)/\sigma$,
i.e. $S/T$ is large near the band edge where the logarithmic derivative
of $\sigma$ with energy is high.

Hicks and Dresselhaus suggested that this conundrum could be overcome
in quantum well structures. \cite{hicks}
They observed that in a two dimensional system the dependence of
the density of states on energy for a parabolic band is a step
function, meaning that for the in-plane direction one
may expect a faster onset of the conductivity with energy
and furthermore that $S$ will be higher for a given
carrier concentration. Viewed in three dimensions, the Fermi 
surfaces of superlattices or two dimensional semiconductors
are in the shape of cylinders or pipes running along
the direction of the layering rather than the spheres or ellipsiods
of three dimensional doped semiconductors.

However, most applications of thermoelectrics involve macroscopic
devices that are difficult to implement with superlattices and
additionally, there can be problems such as parasitic heat conduction
in barrier layers of superlattices.
Nonetheless, it is interesting to observe that Na$_x$CoO$_2$, 
which is representative of the highest performance oxide thermoelectrics,
and has high $ZT$ at high carrier concentration, \cite{terasaki}
has a very two dimensional electronic structure. \cite{singh-nac}
This material illustrates another problem with using 2D electronic
systems as thermoelectrics. The high electrical conductivity is
realized only in the layers, not perpendicular to them, while the
heat conduction is more isotropic. As such, the very high $ZT$ is
realized only in single crystals for in-plane conduction or at
least in highly textured ceramic. Here we propose an alternate
resolution of the conundrum of high $\sigma$ and high $S$ using
low dimensional electronic structures.

We observe that it is possible to have an electronic structure
that is low dimensional in a material that is not low dimensional
provided that symmetry is obeyed. This is known in metallic
materials, the best example being body centered cubic Cr metal,
where flat (i.e. 1D) parts of the Fermi surface give rise to 
an nesting induced spin density wave. \cite{fawcett}
Another example is the superconductor Sr$_2$RuO$_4$,
which dispite its tetragonal symmetry has flat one dimensional
sheets of Fermi surface that give rise to nesting induced
peaks in its susceptibility. \cite{mazin-99,sidis}
Generally, these cases are large Fermi surface metals, which
are not of interest as thermoelectrics.
However, there is no symmetry or other fundamental reason that
this must always be the case and we begin by pointing out
counterexamples.

The face centered cubic rocksalt structure
chalcogenides, PbTe, PbSe, PbS and SnTe are the basis
of excellent thermoelectric materials. \cite{ioffe,wood}
While the thermoelectric properties of these materials has
been discussed in terms of various physical models, band
structure calculations in combination with standard
Boltzmann transport theory can reproduce and predict their
thermopowers, as illustrated by predictions
for PbSe. \cite{parker,zhang-pbse}
As is well known, the valence band ($p$-type) electronic
structure is dominated by $L$-point hole pockets for low
carrier concentrations and $T$,
while at higher carrier concentrations and $T$
transport and other data imply additional electronic features,
often discussed as a second heavy band. \cite{ravich1,ravich2,ekuma}
According to band structure calculations, there is no second heavy
band, but instead connections develop between the $L$-point pockets
near but not at the valence band maximum.

\begin{figure}
\includegraphics[width=0.9\columnwidth]{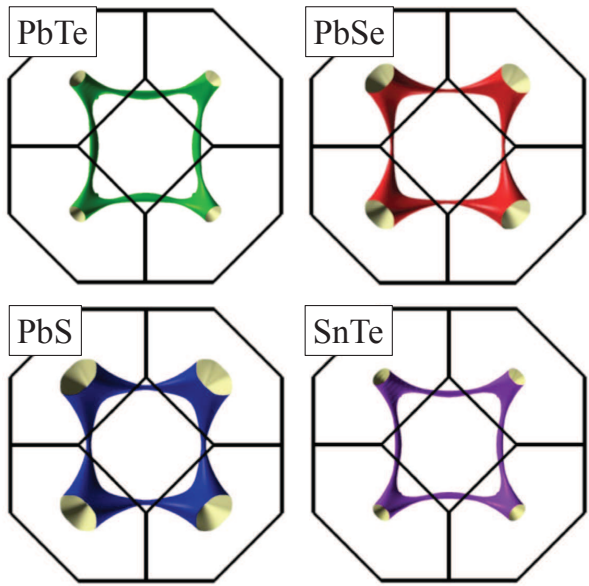}
\caption{(color online)
Calculated valence band constant energy surfaces of
PbTe, PbSe, PbS and SnTe,
at  0.25 eV, 0.49 eV, 0.61 eV, and 0.41 eV
below the valence band maximum, respectively. The corresponding carrier
concentrations in holes per unit cell are 0.005, 0.030, 0.054 and 0.016,
respectively.}
\label{isosurface}
\end{figure}

This is illustrated in Fig. \ref{isosurface}, which
shows energy isosurfaces for the near valence band edge of
PbTe, PbSe, PbS and SnTe. These are based on calculations
done with the augmented planewave plus local orbital
method, \cite{sjo} as implemented in the WIEN2k code. \cite{wien2k}
The calculations included spin-orbit, which is needed for these
materials. We employed the modified Becke-Johnson potential of
Tran and Blaha (TB-mBJ), \cite{tran}
which generally gives improved band
gaps for simple semiconductors and insulators.
\cite{tran,singh-2,koller}
Besides these details the calculations are similar to those
presented previously.
\cite{lin,Wei1997,bilc,Hummer2007,zhang,singh-pbte,parker,xu,svane,singh-fml}
The densities of states (not shown) show low values characteristic
of a light band up to the energy where the $L$-point pockets connect,
at which point there is a sharp onset of a steeply rising density of
states, which is clearly beneficial for obtaining enhanced $S(T)$
at doping levels near the onset and was discussed in the context of the
thermoelectric performance of PbTe. \cite{singh-pbte}
Here we associate this with the pipes.

Qualitatively, the Fermi surface of a doped superlattice or other
2D semiconductor is cylindrical running along the stacking direction.
The conductivity is low along the cylinder and high in the plane.
Considering for example the conductivity along $x$ for a cubic network
of pipes running along $k_x$, $k_y$ and $k_z$ as is approximately the case
in these materials, the pipes along $k_y$ and $k_z$ will contribute as
in a superlattice material in plane, while the pipes along $k_x$ will behave
like the stacking direction and will not contribute to the conductivity.
Thus the energy dependence and other behavior will be the same as the
superlattice, including the enhanced 2D behavior of the thermopower,
except that now the properties will be isotropic due to the cubic
symmetry and superposition of pipes on different directions.

Clearly, the electronic structures
of the chalcogenides shown in Fig. \ref{isosurface} are approximations
of this idealized behavior.   
Nonetheless, they suggest elucidation of
the behavior of a cubic or other three
dimensional semiconductor with a low dimensional
electronic structure in the sense discussed above.
We suggest that this may be a useful paradigm in the search for
new high performance thermoelectric materials.

\section{Calculations}

To describe the behavior of the transport in the aforementioned ``pipes" scenario, we pursue some calculations to elucidate the relevant physics.  We consider the case of a one band material with a pipe-like electronic structure.
We begin with the assumption that the electronic scattering time $\tau(E)$ is independent of energy, i.e. the ``constant scattering time approximation" (CSTA).  The CSTA has been used with quantitative accuracy to describe the thermopower of a substantial number of thermoelectric materials \cite{madsen,singh-pbte,zhang3,ong,singh3,scheide,bertini,lykke,wang} 
so that its usage is on solid practical grounds.  

Then we have the canonical expressions for the electrical conductivity $\sigma(T)$ and Seebeck coefficient $S(T)$:
\bea
\sigma(E) &=& N(E)v^2(E)\tau(E)\\
\sigma(T) &=& -\int_{-\infty}^{\infty} \,\,dE \sigma(E) df(E-\mu)/dE \\
S(T) &=& -\frac{k_{B}}{e\sigma(T)}\int_{-\infty}^{\infty} \,\,dE \sigma(E)\frac{E-\mu}{T} df(E-\mu)/dE
\eea
where $f$ is the Fermi function, $e$ the electronic charge, $k_{B}$ is Boltzmann's constant, $\tau(E)$ is the scattering time, $v(E)$ the Fermi velocity, $\mu$ the chemical potential and $N(E)$ the density of states.  The tensor indices are suppressed for clarity, and in addition the integrations in actual calculations involve a Brillouin-zone sum.

We now compare the thermopower and power factor $S^{2}\sigma$ of two specific idealized Fermi surface topologies: a two dimensional cylindrical Fermi surface connecting the L-points of the fcc Brillouin zone, as suggested by Figure 1, and a three dimensional spherical Fermi surface.  This is an idealization, as in an actual
material Fermi surfaces which contact Brillouin zone faces must do so at
perpendicular angles, so that the pipes must reconnect at the L-point
pockets, as they do in, for example, band structure calculations for PbTe.  Both bands are assumed to be parabolic, and to ensure a fair comparison we choose the radial mass of the cylinder and of the sphere to be equal.  Additionally, as was noted by Ref. \onlinecite{snyder_nature}, in the chalcogenides the cylindrical band is twelve-fold degenerate and we have assumed this here.  For comparison purposes we have assumed the spherical Fermi surface to also be twelve-fold degenerate.  

Then within the CSTA the above integrals are easily evaluated for both the 3D and 2D cases, yielding the following expressions (here $\eta=\mu/T$, the reduced chemical potential.)
\bea
S_{3D}(T) &=& \frac{5}{3}\frac{F_{3/2}(\eta)}{F_{1/2}(\eta)} - \eta \\
\sigma_{3D}(T)& = &\frac{p e^2 \tau}{m^*} \\
S_{2D}(T) &=& 2\frac{F_{1}(\eta)}{F_{0}(\eta)} - \eta \\
\sigma_{2D}(T) & = & \frac{2p e^2 \tau}{3m^*}
\eea
Here $p$ is the carrier density given as
\bea
p &=& \int dE N(E) f(E-\mu)
\eea
where $N(E)$ is the density of states, m$^{*}$ the carrier effective mass, and F is the Fermi-Dirac integral, defined as
\bea
F_{i}(\eta) &=& \int_0^{\infty} x^{i}/(\exp(x-\eta)+1)
\eea
The 2/3 factor for the two-dimensional conductivity arises because, each of the cylinders contributing to the density-of-states, and hence
$p$, conducts in only two of three directions.  The final piece of information necessary to calculate the thermopower is the relation of the reduced chemical 
potential $\eta$ to the carrier concentration $p$, which is done (for each of the two cases) by inverting Eq. 8; details of this procedure can be found in Ref. \onlinecite{blakemore}.

With these mathematical preliminaries completed, we now move to the calculated results. For concreteness (although the results do not sensitively depend
on these assumptions) we have assumed an fcc cell of lattice constant 6.46 \AA, band masses of 0.2 m$_{0}$, where m$_{0}$ is the free electron mass, and fixed the temperature at
1000 K.  This is the approximate maximum operating temperature of the chalcogenides.  For the electrical conductivity we have assumed a doping independent scattering time $\tau$ of 10$^{-15}$ sec, which yields high temperature conductivities of 100 - 1000 $(\Omega$-cm)$^{-1}$, in line with experimental results on these materials.  In Figure 2 we present the calculated thermopower results for the two scenarios.
\begin{figure}
\includegraphics[width=8cm]{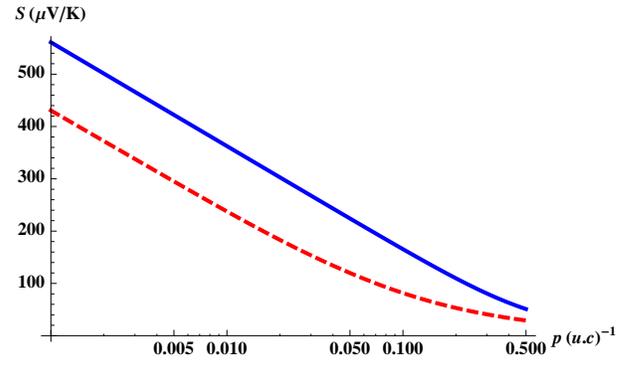}
\caption{(color online) The calculated thermopower for the 2D (blue solid line) and 3D (red dashed line) cases.  Carrier concentrations are given in carriers per unit cell.}
\end{figure}
As is evident, the 2D thermopower exceeds the 3D values by a substantial margin throughout the entire range from 0.001 - 0.5 holes/unit cell.  At the heavy dopings of 0.05 - 0.1 per unit cell, the
2D thermopower is nearly double the 3D value, which is highly favorable for thermoelectric performance, and this thermopower increase comes at a conductivity reduction (Eqs. 5 and 7), relative to the 3D case, of only one third.  Given this, it is not surprising that the calculated 2D power factor (Figure 3) exceeds that of the 3D case across the entire range of concentration modeled, and its maximum value is two and a half times the corresponding 3D maximum.  It is highly likely that correspondingly higher 2D performance (i.e. ZT), relative to the 3D case, would also occur.  We emphasize that the notation 2D and 3D is to distinguish the cases, but that in both cases we are referring to the bulk, macroscopic measurable values for the cubic crystal.
\begin{figure}
\includegraphics[width=8cm]{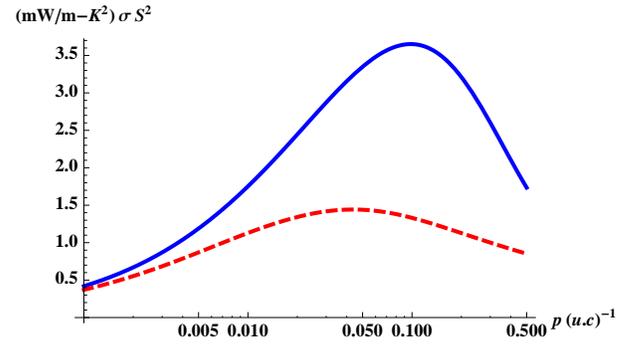}
\caption{(color online) The calculated power factor $S^{2}\sigma$ in mW/m-K$^{2}$ for the 2D (blue solid line) and 3D (red dashed line) cases.}
\end{figure}

\section{Analysis of enhanced Seebeck coefficient in 2D case}

The results of the previous section strongly suggest that the two dimensional ``pipe" topology is favorable for thermoelectric performance, particularly for the Seebeck coefficient, an indispensable ingredient of good thermoelectric performance.  Here we provide analytic understanding of this result.

To a first approximation, the enhanced behavior of the 2D system modeled here can be traced to the relatively larger Fermi surface volume (or, equivalently, carrier concentration) of a 2D cylinder relative to a 3D sphere, for given Fermi energy.  In this case the Fermi surface volume of the cylinder is proportional to the length of the cylinder $=\frac{2\pi}{a}$, which is a value much larger than the radius of the cylinder or the sphere, so that for given Fermi energy the carrier concentration is much larger.  The Fermi energy is relevant because of the well-known Mott formula for the thermopower,
\bea
S &=& \frac{\pi^2 k_B}{3e}k_B Td\log(\sigma(E))/dE_{| E=E_{F}}
\eea
and for a parabolic 3D band yields
\bea
S &=&  \frac{\pi^2 k_B}{2e}k_B T/E_{F}
\eea
so that the thermopower is inversely proportional to the Fermi energy.  In two dimensions, at fixed carrier concentration the Fermi energy is much smaller than in three dimensions, and the thermopower is enhanced as a result.

To gain additional insight into this phenomenon, as well as explore the effect of changing parameters such as the effective mass and temperature, we now pursue analytic calculations within two well-known limits for which closed form results are available: the degenerate limit, when $\eta \equiv E_{F}/T  \gg 1$, and the non-degenerate limit, when $\eta < 0$.  High thermoelectric performance is typically found somewhat between these two regimes, but together these regimes account for most of the behavior of the thermopower in Figure 1.  We begin with the degenerate limit.  In two dimensions, for radial mass m$^{*}$, using the Mott formula it is easy to show (assuming a band degeneracy of 24, 2 for spin and 12 for the 12 ``pipes") that the thermopower takes the form
\bea
S_{2D} &=& \frac{\pi^2}{3}\frac{k_{B}}{e}\frac{3m^*a^2k_{B}}{\pi \hbar^2 p}
\eea
where p is the carrier concentration per unit cell, and similarly for 3D (assuming the same band degeneracy)
\bea
S_{3D} &=& \frac{\pi^2}{2}\frac{k_{B}}{e}\frac{2m^*a^2k_BT}{\hbar^2 (\pi^2 p)^\frac{2}{3}}
\eea
so that one finds the simple result that
\bea
S_{2D}/S_{3D} &=& (\pi/p)^\frac{1}{3}
\eea
Since the carrier concentration $p$ per unit cell is typically much less than unity, one finds S$_{2D}$ to be substantially larger than S$_{3D}$.  Note also that for large unit cells, one requires proportionately greater carrier concentrations {\it per cell} to keep the same carrier concentration {\it per cubic centimeter}, so that for larger cells the above equation has a smaller range of validity.  Numerically for p=0.5/u.c. (yielding an $\eta_{2D}$ of 5.5) this ratio is 1.845, while the exact result is 1.747, less than a 6 percent difference.

To treat the non-degenerate limit, we note that in this limit thermopower is generally logarithmic in carrier concentration (the straight lines in Figure 1), so it is the prefactors that are at issue.  The non-degenerate limit is typically specified by $\eta \ll 0$, so that the Fermi function $(\exp(\frac{E-\mu}{T})+1)^{-1}$ reduces simply to $\exp\frac{\mu-E}{T}$ and the integrals involving the Fermi function can be done exactly.  In this limit, as is well known \cite{goldsmid}, the three dimensional parabolic band thermopower is given by
\bea
S(p,T)_{3D} &=& \frac{k_B}{e}(\frac{5}{2} - \eta_{3D}(p,T))
\eea
It is a simple matter to work out the corresponding value for our 2D cylindrical parabolic band and one finds that
\bea
S(p,T)_{2D} &=& \frac{k_B}{e}(2 - \eta_{2D}(p,T))
\eea
Note that $\eta_{2D}$ and $\eta_{3D}$ vary due to the topology difference, and we now work out an expression for their difference.  For 2 dimensions, the relation of 
$\eta$ and p can be evaluated easily and is simply (for all temperatures)
\bea
\eta_{2D} &=& \log(\exp(\frac{\pi p}{3m^*Ta^2})-1)
\eea
and in the non-degenerate limit this becomes simply
\bea
\eta_{2D} &=& \log(\frac{\pi p}{3m^*T a^2})
\eea
One can similarly work out an expression for $\eta_{3D}$ in the non-degenerate limit and one finds
\bea
\eta_{3D} &=& \log\left(\frac{4 \pi^{3/2} p}{3 (2m^*)^{\frac{3}{2}}a^3T^{\frac{3}{2}}}\right)
\eea
so that, restoring the appropriate powers of $\hbar$ and k$_{B}$  one finds that
\bea
\eta_{2D}-\eta_{3D} = - \log\left(\frac{m^{*^{1/2}}a (k_{B}T)^{\frac{1}{2}}}{\sqrt{2\pi} \hbar}\right)
\eea
For the modeled situation ($m^*=0.2 m_{0}, T=1000 K, a= 6.46 $\AA) the difference is -2.097 so that 
in the non-degenerate limit one finds $S_{2D}-S_{3D}=1.597 k_{B}/e= 137 \mu V/K$, which is very close to the difference in these values
at the left hand of Figure 1.  This is a substantial increase, needless to say.

The last equation reveals that if the effective mass (which was chosen on the basis of effective masses in the chalcogenides and Bi$_2$Te$_3$) is larger, the effective benefit in the non-degenerate limit is smaller,  but for large effective mass materials one is typically closer to the degenerate limit.  Conversely, if the temperature is smaller (such as for room temperature applications) the difference is correspondingly greater, provided the sample remains in the non-degenerate limit.

\section{Summary and Conclusions}

To summarize, we have here shown that (1) low dimensional electronic structures can occur even in cubic semiconductors, and that (2) such electronic structures are highly beneficial for thermoelectric performance.  This represents a new paradigm for high thermoelectric performance: {\it low-dimensional} electronic structures enhancing performance in fully {\it three dimensional bulk} thermoelectrics.
We suggest searching for new thermoelectric materials among such compounds.  One such compound may be SnTe \cite{littlewood, tanaka}.

{\bf Acknowledgment}

This research was supported by the U.S. Department of Energy, EERE, Vehicle Technologies, Propulsion Materials Program (DP), the ÔSolid State Solar-Thermal Energy Conversion Center (S3 TEC)Õ, an Energy Frontier Research Center funded by the US Department of Energy, Office of Science, Office of Basic Energy Sciences under Award Number: DE-SC0001299/DE-FG02-09ER46577 (XC,DJS).

\end{document}